\documentclass[aps,prb,preprint]{revtex4-1}

\usepackage{graphicx}
\usepackage{dcolumn}
\usepackage{amsfonts}

\begin{document}

\title{Selective photoexcitation of exciton-polariton vortices}
\author{Ga\"el Nardin, Konstantinos G. Lagoudakis, Barbara Pietka, Fran\c cois Morier-Genoud, Yoan L\'eger, Beno\^it Deveaud-Pl\'edran}
\affiliation{Laboratoire d'Opto{\'e}lectronique Quantique, {\'E}cole Polytechnique F{\'e}d{\'e}rale de Lausanne (EPFL), Station 3, CH-1015 Lausanne,
Switzerland.}

\begin{abstract}
We resonantly excite exciton-polariton states confined in cylindrical traps. Using a homodyne detection setup, we are able to image the phase and amplitude of the confined polariton states. We evidence the excitation of vortex states, carrying an integer angular orbital momentum $m$, analogous to the transverse $TEM_{01^{*}}$ ``donut'' mode of cylindrically symmetric optical resonators. Tuning the excitation conditions allows us to select the charge of the vortex. In this way, the injection of singly charged ($m=1$ $\&$ $m=-1$) and doubly charged ($m=2$) polariton vortices is shown. This work demonstrates the potential of in-plane confinement coupled with selective excitation for the topological tailoring of polariton wavefunctions.
\end{abstract}

\maketitle

Exciton-polaritons are hybrid light-matter quasi-particles arising from the strong coupling of the optical mode of a semiconductor microcavity with the excitonic resonance of a quantum well (QW) \cite{Weisbuch}. Several recent proposals have highlighted the potential of tailoring polariton wavefunctions by lateral in-plane confinement for applications in optelectronics devices and quantum information processing \cite{Liew,Carusotto,Johne}. In this paper, we experimentally demonstrate that by combining specific in-plane confinement with selective resonant excitation, we can realize in a controlled way polariton quantized vortices. In opposition to the vortices that spontaneously appear in the Bose-condensed phase of  quasi-2D polaritons \cite{Lagoudakis}, our polaritons vortices are the result of optical vortices strongly coupled with the excitonic resonance. Generation of optical vortices has been demonstrated in different types of lasers like ring resonators \cite{Brambilla}, microchip solid state lasers \cite{Chen} and Vertical Cavity Surface Emitting Lasers (VCSELs) \cite{Colstoun}. They have been identified as the transverse modes of cylindrically symmetric optical resonators and often called the $TEM_{01^{*}}$ ``donut'' mode, which can be obtained as a superposition of the $TEM_{10}$ and $TEM_{01}$ modes \cite{Epler,Degen}.

Cylindrically symmetric polariton traps can be obtained through lateral in-plane confinement by etching micropillars \cite{Bloch} or patterning mesas on the microcavity spacer \cite{ElDaif}. In both types of structure, strong coupling between the confined electromagnetic modes of the cylindrical resonator with the excitonic resonance has been observed, demonstrating the creation of 0D polaritons. Although the confinement potential is acting on the photonic part of the polariton, the strong coupling regime results in the confinement of the whole polariton state, including the excitonic part \cite{Kaitouni}.

\begin{figure}[ht]
	\centering
		\includegraphics{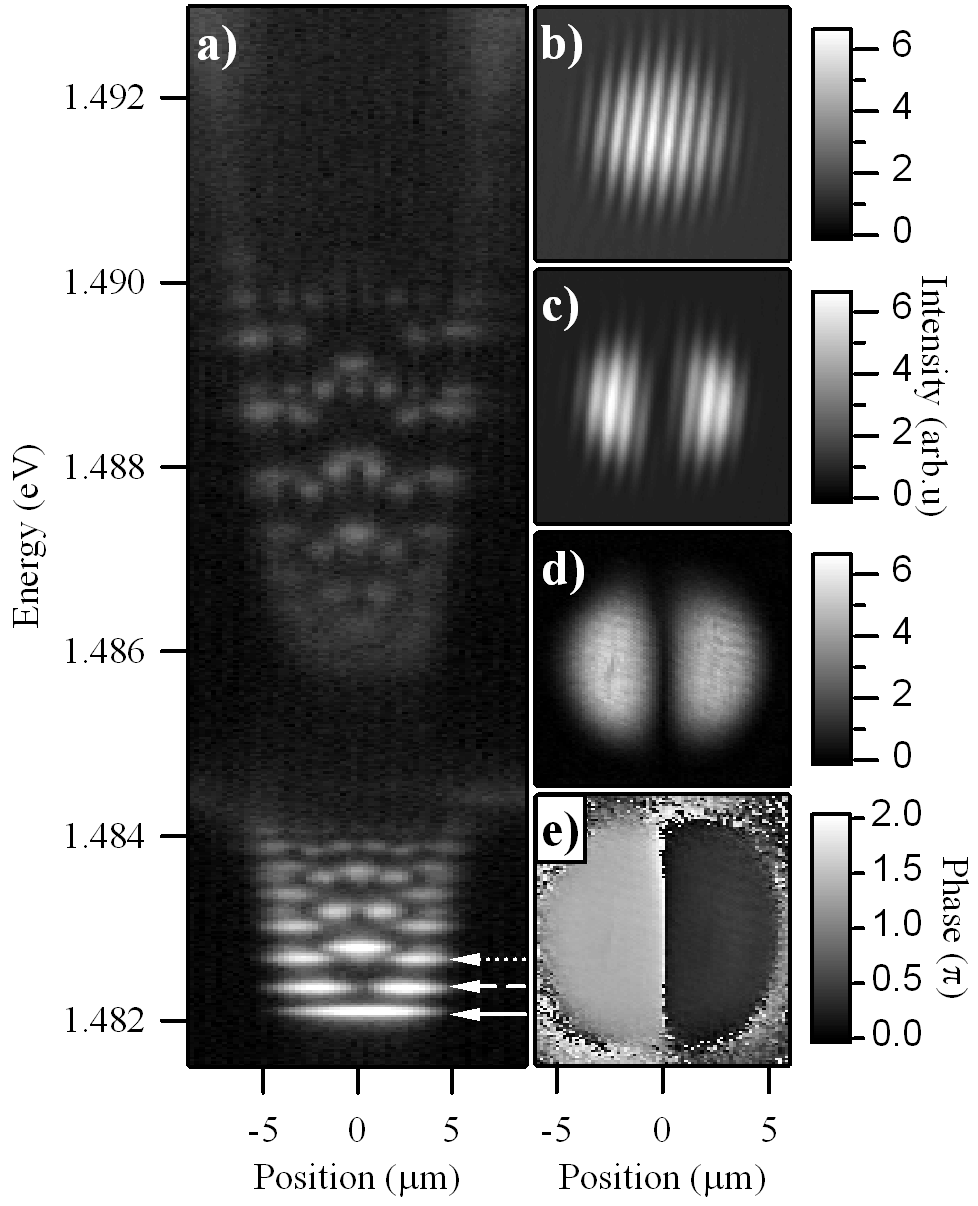}
	\caption{(a) Photoluminescence intensity emitted by the confined polariton states in a trap of $10\mu m$ in diameter, under non-resonant pumping (at 1.57 eV, in the first minimum of reflection of the DBR), in a logarithmic scale. Confined lower (upper) polariton states are visible below (above) 1.484 eV. Solid, dashed and dotted arrows indicate the lower polariton ground, first and second excited states respectively. (b) Interferogram resulting from the interference of the coherent emission of the lower polariton ground state ($n=1,m=0$) with the reference laser. (c),(d) and (e) are respectively the interferogram, the wavefunction amplitude and wavefunction phase of the lower polariton first excited doublet state ($n=1,m=\pm1$).}
	\label{Figure1}
\end{figure}

Our sample is a patterned GaAs $\lambda-$cavity with one embedded InGaAs QW, sandwiched between two semiconductor distributed bragg reflectors (DBRs), featuring a vacuum Rabi splitting of $3.5$ $meV$. The polariton traps consist of circular mesas that were etched on the microcavity spacer. Discrete confined polariton states have been observed in traps with diameters varying between $3\mu m$ and $20\mu m$ \cite{Kaitouni,Nardin0}. A spatially resolved spectrum of the photoluminescence emitted by the confined states in a $10\mu m$ diameter trap under non-resonant pumping is shown in Fig. \ref{Figure1} (a). Discrete eigenstates can be observed for the lower polariton branch (below $1.484 eV$) and the upper polariton branch (above $1.484 eV$). The linewidth of these states is of the order of $80 \mu eV$. All further measurements presented in the paper were performed on lower polariton states confined in mesas of $10\mu m$ diameter, and for a detuning of $\delta\sim0$ $meV$ between the confined photonic mode and the excitonic resonance ($1.484eV$). Similar results were obtained on the confined UP branch.

Imaging wavefunctions of confined polaritons is possible by means of optical microscopy. Indeed, due to their very small effective mass (four orders of magnitude smaller than the free electron mass), polaritons can be confined in traps of sizes in the micrometer range, above the optical resolution limit. Images of the two-dimensional distribution of the confined state probability densities can be realized either using a tomography technique \cite{Nardin1}, or by directly imaging the coherent emission of a given state when it is resonantly excited with a continuous wave laser \cite{Cerna}. The emission intensity gives direct information on the probability density of the polariton states. 

To image the wavefunction rather than the probability density only, one needs a phase-resolved detection scheme. In this perspective, we used a homodyne detection setup, where we split the $cw$ pump laser into two. One part was focalized on the back side of the sample using a camera objective to resonantly excite the polariton states. The camera objective provided us a diffraction limited gaussian spot of $\sim 15\mu m$ diameter (of the same order of magnitude than the size of the mesa) and allowed for a good control of the excitation angle. Both these features are important for the selective excitation of polariton states, as it will be discussed further. The other part of the laser served as a phase reference. The sample was held in a cold-finger cryostat at a temperature of $\sim 4K$. The coherent emission of the polariton state was collected from the front side using a $0.5$ N.A. microscope objective. The image of the emission was then interferred with the reference beam on a CCD. Figure \ref{Figure1} (b) shows the emission of the ground state $(n=1,m=0)$ of the lower confined polariton [indicated with a plain arrow in Fig. \ref{Figure1} (a)] interfering with the reference laser. A slightly different incidence angle is used for the signal and the reference to provide straight interference fringes. As the ground state is expected to have a constant phase, the phase gradient obtained with this state is used as a phase reference for the other interferograms. Using numerical Fourier transform and digital off-axis filtering\cite{Cuche}, one can then extract the amplitude and phase of the polariton states. 

In circular coordinates, confined states can be described by two numbers $(n,m)$, the well known radial and orbital quantum numbers. In cylindrically symmetric systems, $+m$ and $-m$ states are degenerate. In most of the mesas, due to a small ellipticity of the trap, there is a small degeneracy lift of the $\pm m$ doublet into two new eigenstates $\frac{1}{\sqrt{2}}\left[\psi_{+m}\pm\psi_{-m}\right]$ \cite{ElDaif,Nardin1,Cerna}. For example, the first excited state characterized by the quantum numbers $(n=1,m=\pm1)$ [indicated with a dashed arrow in Fig. \ref{Figure1} (a)] is split into two eigenstates that are similar to the $TEM_{01}$ and $TEM_{10}$ modes of transverse laser patterns. The interferogram of one of these eigenstates is shown in Fig. \ref{Figure1} (c). In order to specifically excite this eigenstate, one need to slightly red-shift the excitation laser with respect to the doublet energy. We have extracted the amplitude [Fig. \ref{Figure1} (d)] and the phase [Fig. \ref{Figure1} (e)] of the polariton wavefunction. In Fig. \ref{Figure1} (e), the $\pi$-phase shift between the two lobes is clearly visible. It indicates that the left lobe of the wavefunction is of opposite sign than the right lobe. Around the trap the phase is not defined, as the wavefunction is exponentially decaying outside the trap. 
The ellipticity being very small, the corresponding splitting is smaller than the linewidth. Therefore, the ellipticity can be overcome by pumping between the energies of the two split states, and excite a combination of them. In this case, the azimuthal dependence of the wavefunction reads as: \begin{center} 
$\psi(\phi)=A\psi_{+m}(\phi)+B\psi_{-m}(\phi)=Ae^{im\phi}+Be^{-im\phi}$\end{center} where $A$ and $B$ are complex coefficients. The value of coefficients $A$ and $B$ is given by the overlap (in real space, reciprocal space and energy) between the $\pm m$ states and the pump. When the laser is focused on the mesa, both $+m$ and $-m$ components are generally excited, creating standing wave patterns, similar to the one observed in Fig. \ref{Figure1} (c). The excitation laser can be used to lock, and subsequently control, the pattern in any arbitrary direction \cite{Cerna}. 

\begin{figure}[ht]
	\centering
		\includegraphics{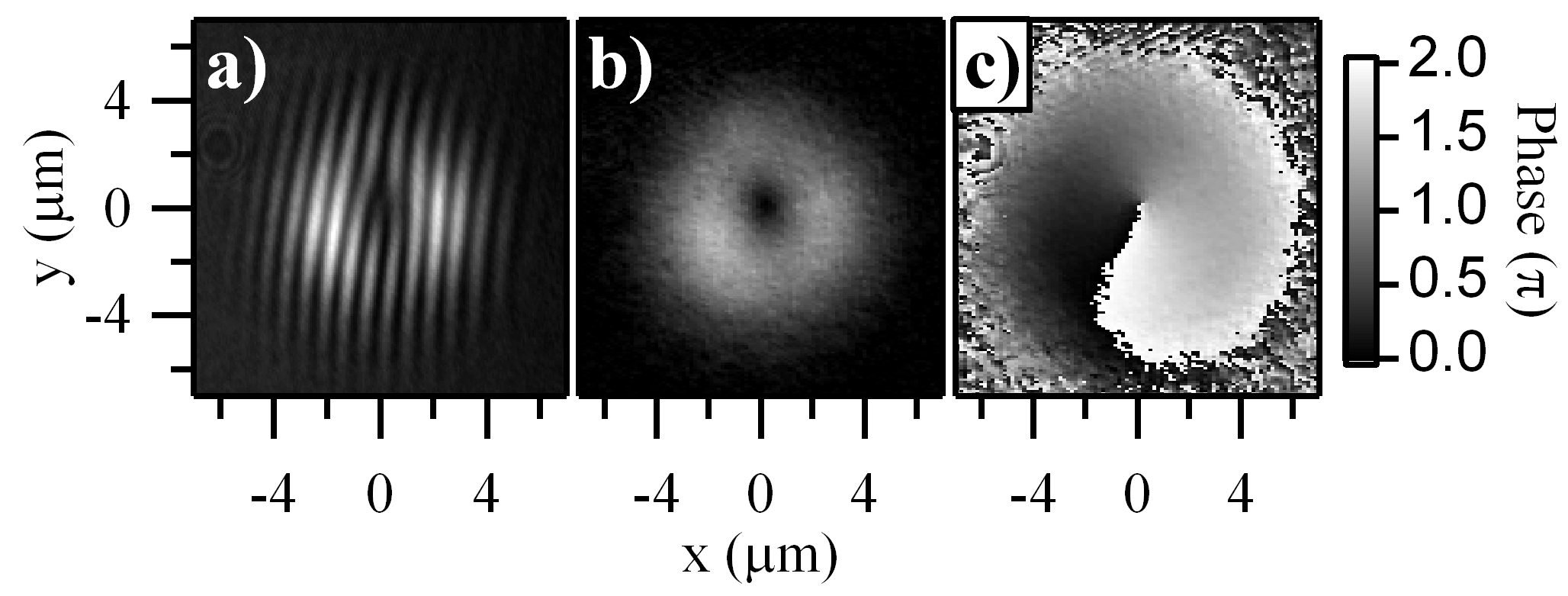}
	\caption{(a) Interferogram resulting from the interference of the coherent emission of the ($n=1,m=+1$) state of the lower polariton with the reference laser, displaying a clear fork-like dislocation. (b) Amplitude of the coherent polariton field, extracted from (a), showing a minimum in the field density at the place of the phase singularity. (d) Phase of the coherent polariton field, extracted from (a), indicating a $2\pi$-phase shift rotation of the phase around the singularity.}
	\label{Figure2}
\end{figure}

We are now going to show that by carefully selecting certain pumping conditions one can create ``pure'' $+m$ or $-m$ states, carrying an integer angular orbital momentum. Focusing the excitation laser on the side of the trap, with a finite excitation angle allows to inject polaritons with a well defined in-plane momentum mainly on one side of the trap. This can be used to select which one of the $+m$ or $-m$ state is injected. Such a state is experimentally produced and shown in Fig. \ref{Figure2} for ($n=1,m=+1$). In the interferogram displayed in Fig. \ref{Figure2} (a), one can observe a very clear fork-like dislocation, indicating the presence of a phase singularity. We extract from this interferogram the amplitude [Fig. \ref{Figure2} (b)] and phase [Figure \ref{Figure2} (c)] of the polariton field. The two characteristics of quantized vortices are observed: the intensity minimum at the center of the trap (b) and the $2\pi$-phase shift around the core (c) are straightforwardly visible. The core of this vortex is situated at the center of the trap, and the size of the vortex is delimited by the mesa diameter. This state is analogous to a $TEM_{01^{*}}$ ``donut'' transverse laser mode.

\begin{figure}[ht]
	\centering
		\includegraphics{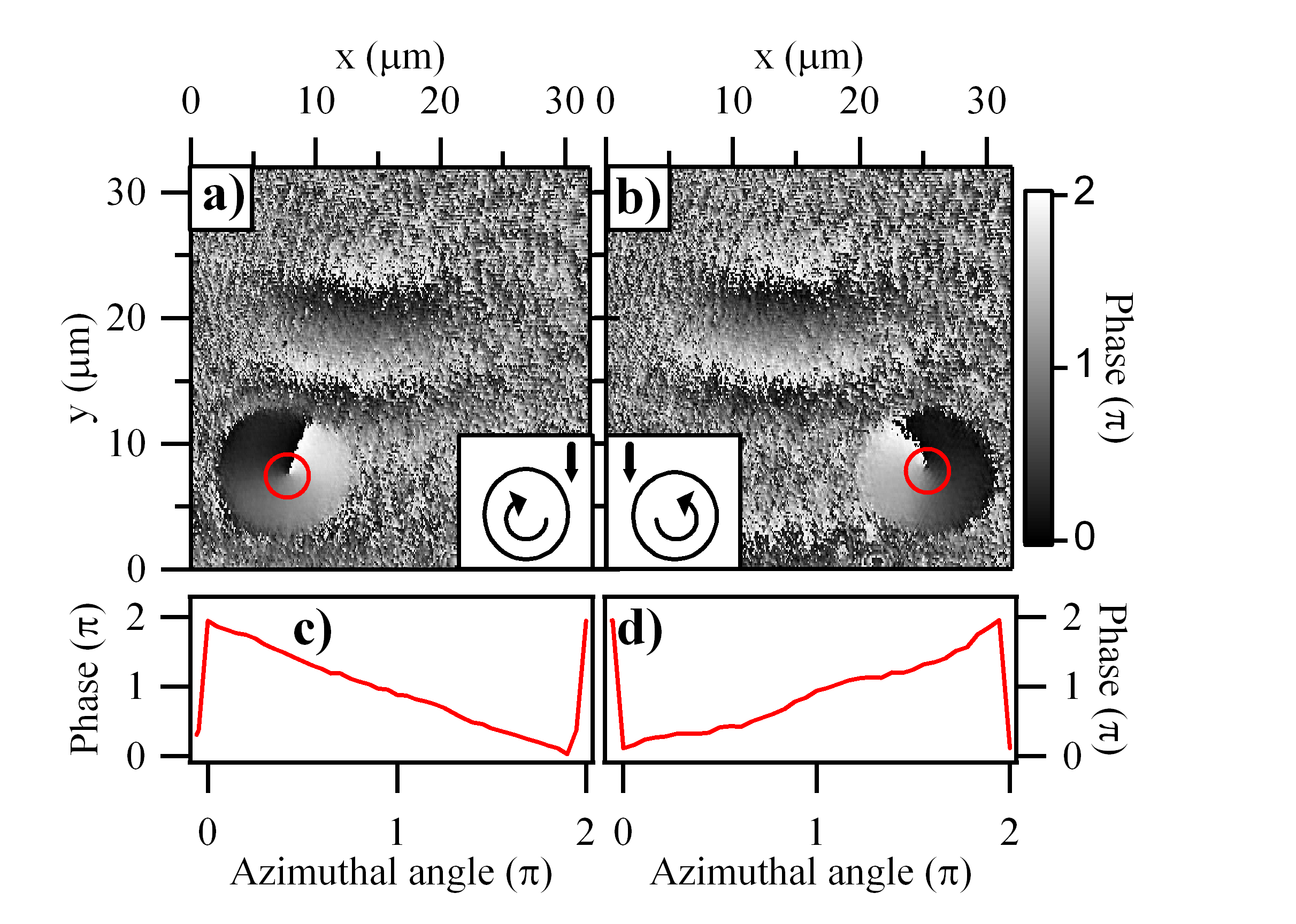}
	\caption{(color online) Phase mapping (a) and phase distribution profile along the red circle (c) of the polaritons in the trap, when the polaritons are injected on the right side of the mesa using an pump angle of around $4^\circ$. In this configuration, the vortex is rotating clockwise. (b-d) Same as (a-c), but injection on the left side of the mesa. In this configuration, the vortex is rotating counterclockwise. Insets: schematic view of the vortex rotation, the pump and its in-plane direction are represented by the thick arrow.}
	\label{Figure3}
\end{figure}

Depending on the side of the trap on which the laser is focused, the charge is going to be positive or negative, as demonstrated in Fig. \ref{Figure3}. The phase distribution [Fig. \ref{Figure3}, (a)-(b)] and the phase profile along the red circle [Fig. \ref{Figure3}, (c)-(d)] are displayed when the mesa is positioned on the left side [Fig. \ref{Figure3} (a)-(c)] or right side [Fig. \ref{Figure3} (b)-(d)] of the excitation spot. It is interesting to note that this selective excitation scheme is compatible with the picture of the injection of a polariton fluid in the trap. The treatment of the polariton field as a propagating fluid has been recently highlighted in Ref. \cite{Amo}.

\begin{figure}[ht]
	\centering
		\includegraphics{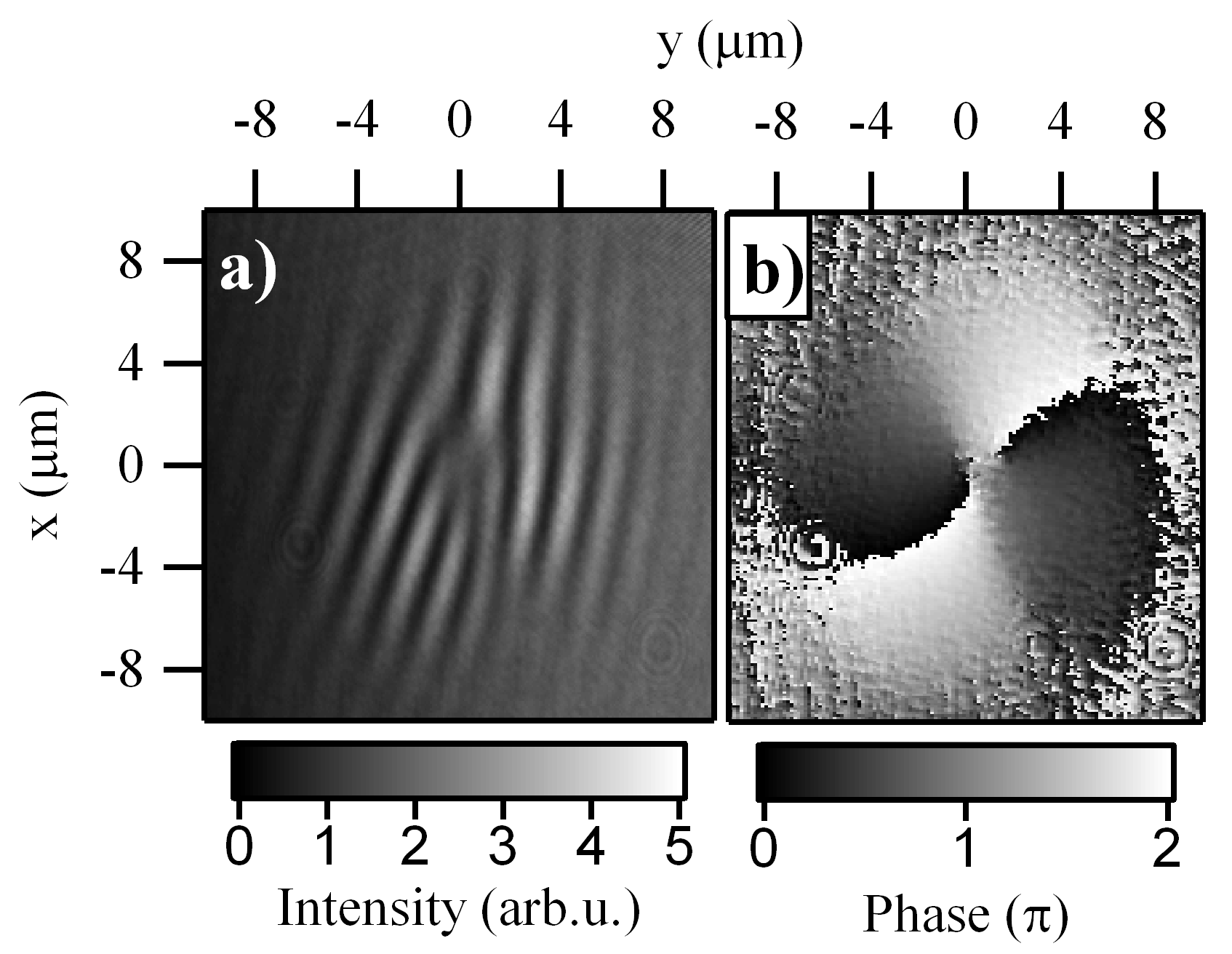}
	\caption{(a) Interferogram resulting from the interference of the coherent emission of the lower polariton second excited state ($n=1,m=+2$) with the reference laser, displaying a trident-like dislocation. (b) Phase mapping extracted from (a), displaying a doubly charged vortex structure.}
	\label{Figure4}
\end{figure}

It is also possible to select the value of the vortex charge, by tuning the excitation energy to be resonant with a state carrying another quantum number $m$. Figure \ref{Figure4} displays the interferogram [Fig. \ref{Figure4} (a)] and corresponding phase mapping [Fig. \ref{Figure4} (b)] of a ($n=1,m=+2$) state [indicated with a dotted arrow in Fig. \ref{Figure1} (a)]. A trident-like dislocation is visible in the interferogram, resulting in a $4\pi$-phase shift around the core. 

All the measurement presented in this paper were performed in the low excitation density regime. Increasing the pump power by three orders of magnitude (up to $3.2 kW/cm^2$) allowed us to observe the effect of polariton-polariton interactions in the form of a blue-shift of the vortex state of the order of $120\mu eV$. A recent publication \cite{Krizhanovskii} investigated the effect of interactions on the size of a polariton vortex imprinted using an optical parametric oscillator (OPO) process, and reported a reduction of the vortex core diameter down to $3 \mu m$ when increasing the excitation density. On the range of excitation power used in our experiment, we never observed any change in the vortex core size. This result is consistent with Reference \cite{Krizhanovskii}, as in our case, the size of the core is already limited to $3\mu m$ by the confining potential, even in the low polariton density regime. Experiments are in progress to track this effect with mesas of different sizes and detunings.

In conclusion, we have demonstrated the selective excitation of trapped polariton states in a patterned semiconductor microcavity. Thanks to the homodyne detection setup, the phase and the amplitude of the coherent polariton gas were directly visualized, and trapped polariton vortices were identified as a superposition of eigenstates of the quasi-circular traps. We have experimentally shown the selection of the vortex charge by tuning the excitation conditions. This work shows the essential role of engineered lateral confinement to tailor the wavefunction and the subsequent emission pattern of a resonantly excited exciton-polariton gas. This will possibly provide new exploitation schemes for polariton lasers which also have a cylindrical symmetry \cite{Bajoni1}.

We would like to thank  T.K Para\"iso, R. Cerna, M.T. Portella-Oberli, O. El Daif, N. Pavillon and B. Caire-Remonnay for helpful discussions. We acknowledge support by the Swiss National Research Foundation through the 'NCCR Quantum Photonics'.

\pagebreak

\pagebreak

\end{document}